\documentclass[a4paper,oneside,11pt]{article}

\newtheorem{dfn}{Definition}[section]
\newtheorem{tw}[dfn]{Theorem}
\newtheorem{prop}[dfn]{Proposition}
\newtheorem{rem}[dfn]{Remark}

\newtheorem{lem}[dfn]{Lemma}

\usepackage{anysize}
\usepackage{verbatim}
\usepackage{array}
\usepackage{enumerate}
\usepackage{amssymb} % symbole AMS
\usepackage{amsmath}
\usepackage{fancyhdr}
\usepackage{euscript}
\usepackage{latexsym}

\author{Micha\l \ Baran \\ \small  Mathematics Department of Cardinal Stefan Wyszy\'nski University in Warsaw, Poland\\ \small{\it m.baran@uksw.edu.pl} \bigskip \\
\\
Jerzy Zabczyk \footnote{Sponsored by the European Transfer of
Knowledge project SPADE2.}\\ \small Institute of Mathematics, Polish
Academy of Sciences,
     Warsaw,  Poland\\ \small{\it zabczyk@impan.pl}}

\title{\bf Bonds with volatilities proportional to  forward rates}
\begin{document}
\baselineskip=1.1\baselineskip \maketitle
\date
\

\begin{abstract}
The problem of existence of solution for the Heath-Jarrow-Morton
equation with linear volatility and purely jump random factor is
studied. Sufficient conditions for existence and non-existence of
the solution in the class of bounded fields are formulated. It is
shown that if the first derivative of the L\'evy-Khinchin exponent
grows slower then logarithmic function then the answer is positive
and if it is bounded from below by a fractional power function of
any positive order then the answer is negative. Numerous examples
including models with  L\'evy measures of stable type are presented.
\end{abstract}

\noindent
\begin{quote}


\end{quote}

\bigskip
\section{Introduction}
We are concerned with the bond market model, on a fixed time
interval $[0, T^*]$, $T^\ast<\infty$, in which the bond prices $P(t,
T),\,\,0\leq t \leq T \leq T^*\,,$ are represented in the form,
\begin{gather*}
P(t,T)= e^{-\int_{t}^{T}f(t,u)du}, \qquad t\leq T\leq T^\ast.
\end{gather*}
Moreover, forward curves processes $f(t,T),\,\,0\leq t \leq T \leq
T^*\,,$ are It\^o processes with stochastic differentials:
\begin{gather}\label{rownanie na f}
df(t,T)=\alpha(t,T)dt+\sigma(t,T)dL(t),\qquad (t,T)\in\mathcal{T},
\end{gather}
where
\begin{gather}\label{dziedzina f}
\mathcal{T}:=\left\{(t,T)\in\mathbb{R}^2: 0\leq t\leq T\leq
T^\ast\right\}.
\end{gather}
The random factor process $L$ is a real L\'evy process defined on a
fixed probability space $(\Omega,\mathcal{F},P)$.

\noindent
One can extend the definition of $f$ given by
\eqref{rownanie na f} on the set $[0,T^\ast]\times[0,T^\ast]$ by
putting
\begin{gather}\label{zerowanie wspolczynnikow}
\alpha(t,T)=0, \qquad\sigma(t,T)=0 \quad \text{for} \
t\in(T,T^\ast].
\end{gather}

\noindent Let $\mathcal{P}$ and $\mathcal{O}$ denote the predictable
and optional $\sigma$- field on $\Omega\times[0,T^\ast]$
respectively. We follow \cite{EbJacRaib} in imposing assumptions on
the drift and volatility coefficients in \eqref{rownanie na f}:
\begin{gather}\label{prognozowalnosc wspolczynnikow}
(\omega,t,T)\longrightarrow \alpha(\omega,t,T),
\sigma(\omega,t,T)\qquad \text{are} \
\mathcal{P}\otimes\mathcal{B}([0,T^\ast]) \
\text{measurable}\\[1ex]\label{ograniczonosc wspolczynnikow}
\sup_{0\leq t,T\leq
T^\ast}\Big\{\mid\alpha(t,T)\mid+\mid\sigma(t,T)\mid\Big\}<\infty.
\end{gather}
Conditions \eqref{zerowanie wspolczynnikow}-\eqref{ograniczonosc
wspolczynnikow} provide that we can find a version of $f$ such that
for each $T\in[0,T^\ast]$
\begin{gather}\label{opcjonalnosc rozwiazania}
(\omega,t,T)\longrightarrow f(\omega,t,T), \qquad t\leq T\leq
T^\ast\quad \text{is} \ \mathcal{O}\otimes\mathcal{B}([0,T^\ast])\
\text{measurable}.
\end{gather}

\bigskip \noindent Condition \eqref{zerowanie wspolczynnikow} implies
that
\begin{gather*}
f(t,T)=f(T,T),\qquad \text{for} \ t\in(T,T^\ast]
\end{gather*}
and consequently that the discounted bond price process defined by
\begin{gather*}
\hat{P}(t,T):=e^{-\int_{0}^{t}r(s)ds}\cdot P(t,T),\qquad (t,T)\in
[0,T^\ast]\times[0,T^\ast],
\end{gather*}
with a short rate $r(t):=f(t,t)$, is given by the formula
\begin{gather*}
\hat{P}(t,T)=e^{-\int_{0}^{T}f(t,u)du}, \qquad (t,T)\in
[0,T^\ast]\times[0,T^\ast].
\end{gather*}

\noindent  If one assumes in addition that $\hat{P}(\cdot,T), \
T\in[0,T^\ast]$ are local martingales then for each
$T\in[0,T^\ast]$,  see \cite{Eberlein}, \cite{JakubowskiZabczyk},
\begin{gather}\label{warunek HJM}
\int_{t}^{T}\alpha(t,u)du=J\left(\int_{t}^{T}\sigma(t,u)du\right)
\end{gather}
for almost all $t\in[0,T]$, where the function $J$ is the L\'evy -
Khinchin exponent determined by the Laplace transform:
\begin{gather*}\label{J}
\mathbf{E}(e^{-zL(t)})=e^{tJ(z)},\qquad t\in[0,T^\ast], \ z\in
\mathbb{R}.
\end{gather*}
Of prime interest is to find out under what conditions one can model
bond prices  with volatility proportional to  forward curves :
\begin{gather}\label{linearity}
\sigma (t, T) =\lambda (t,T)  f(t-,T), \qquad (t,T) \in \mathcal{T},
\end{gather}
where $\lambda$ is a  continuous deterministic function on
$\mathcal{T}$ bounded from below and from above by positive
constants $\underline {\lambda}$, $\bar {\lambda}$:
$$
0< \underline {\lambda}\,\,\leq \,\,\lambda (t,T) \,\,\leq \,\,  \bar{\lambda} < +\infty,\,\,\,\,(t,T)\in\mathcal{T}.
$$
Obviously one can choose $\bar{\lambda}$ arbitrarily large. For
technical reasons we assume that $\bar{\lambda}\geq 1$.

\vspace{2mm} {\it This problem has been first stated in
\cite{Morton} in the case when $L$ is a Wiener process and solved
with a negative answer: linearity of volatility implies explosion of
forward rates, see \cite{Morton} Section 4.7 or \cite{Filipovic} ,
Section 7.4. This fact was one of the main reasons that the BGM
model was formulated  in terms of Libor rates and not in terms of
forward curves, see \cite{BGM}}.\vspace{5mm}

Differentiating the identity \eqref{warunek HJM} with respect to $T$
and taking into account the condition (\ref{linearity}) we see that
proportionality of the volatility implies that the forward curve
satisfies the following equation on $\mathcal{T}$,
\begin{gather}\label{f rownanie}
df(t,T)=J^{'} \left(\int_{t}^{T}\lambda (t,u)
f(t-,u)du\right)\lambda (t,T) f(t-,T) dt+ \lambda(t,T)
f(t-,T)dL(t)\,\,.
\end{gather}
with initial condition,
\begin{gather}\label{initial}
f(0,T) = f_{0} (T),\,\,\,\,T \in [0, T^{*}].
\end{gather}
In particular if $L$ is a Wiener process then $J(z)=\frac{1}{2}z^2$
and if $\sigma(t,T)=f(t,T)$ then \eqref{f rownanie} becomes
\begin{gather*}
df(t,T)= \left(\int_{t}^{T} f(t,u)du\right)f(t,T) dt+
f(t,T)dL(t),\qquad (t,T)\in \mathcal{T}.
\end{gather*}
This equation has been studied in \cite{Morton}.

\noindent
Taking into account \eqref{zerowanie
wspolczynnikow}-\eqref{opcjonalnosc rozwiazania} we assume that
\begin{gather*}
\lambda(t,T)=0\qquad\qquad \text{for} \ t\in(T,T^\ast],
\end{gather*}
and we search for a solution $f$ of \eqref{f rownanie} in the class
of random fields satisfying the following conditions
\begin{gather}\label{1war na rozw}
(\omega,t,T)\longrightarrow f(\omega,t,T),\quad 0\leq t\leq T\leq
T^\ast \ \text{is} \ \mathcal{O}\times\mathcal{B}([0,T^\ast])\
\text{measurable},\\[1ex]\label{2war na rozw}
f(\cdot,T) \ \text{is c\`adl\`ag on} \ [0,T] \
\text{for each} \ T\in[0,T^\ast]\\[1ex]\label{3war na rozw}
(\omega,t,T)\longrightarrow f(\omega,t-,T)\
\text{is} \ \mathcal{P}\times\mathcal{B}([0,T^\ast])\
\text{measurable},\\[1ex]\label{4war na rozw}
\sup_{(t,T)\in\mathcal{T}}f(t,T)<\infty, \qquad P-\text{a.s.}.
 \end{gather}
Requirement \eqref{4war na rozw} states that the function $f(\omega,
\cdot,\cdot)$ is bounded on $\mathcal{T}$ but notice that the bounds
may depend on $\omega$. Random fields satisfying \eqref{1war na
rozw}-\eqref{4war na rozw} will be called the class of {\bf bounded
fields} on $\mathcal{T}$.

\noindent We also examine explosions of solutions from the class of
locally bounded fields. For $0<x\leq T^\ast$, $0<y\leq T^\ast$
consider a family of subsets of $\mathcal{T}$ given by
\begin{align}\label{def zbioru T_x,y}
\mathcal{T}_{x,y}:=\left\{(t,T)\in\mathcal{T}:0\leq t\leq x,0\leq
T\leq y \right\}.
\end{align}
A random field is {\bf locally bounded} if it is bounded on
$\mathcal{T}_{T^\ast-\delta,T^\ast-\delta}$ for each
$0<\delta<T^\ast$.

The main question of the paper is concerned with existence or
non-existence of solutions to \eqref{f rownanie} - \eqref{initial}.
We derive conditions on the L\'evy process $L$ under which there
exists a bounded field solving \eqref{f rownanie}, see Theorem
\ref{tw 2 glowne} and conditions under which such solutions do not
exist, see Theorem \ref{tw 1 glowne}. In the latter case we assume
that $\lambda$ is equal to $1$. Under assumptions of Theorem \ref{tw
1 glowne} we also show that if there exists a locally bounded field
$f$ solving \eqref{f rownanie} then it {\bf explodes}, i.e.
\begin{gather*}
\lim_{(t,T)\rightarrow(T^\ast,T^\ast)}f(t,T)=+\infty,
\end{gather*}
see, Theorem \ref{tw1 o eksplozjach }. From general
characterizations explicit conditions on the jumps of the random
factor are deduced implying existence or non-existence of models
with proportional volatilities. Results for models with negative
jumps are stated as Theorem \ref{tw 3} and Theorem \ref{przyklad 2}
and with strictly positive jumps in Theorem \ref{przyklad6}, Theorem
\ref{przyklad4} and Theorem \ref{large jumps}. Note that models with
positive jumps are very attractive from the practical point of view.
In fact typical shocks shift forward curves upwards what is
equivalent to drops in bond prices. Special cases of our existence
results can be deduced, via Musiela parametrization, from results
presented in \cite{PeszatZabczyk}. The method of establishing the
results on non-existence was inspired by the idea of Morton in
\cite{Morton}, where the solution is being compared with a
deterministic exploding function. \vspace{3mm}

\noindent The paper is organized as follows. Section
\ref{Preliminaries} contains preliminaries necessary to formulation
and proofs of the main results of the paper. Section \ref{Main
results} is devoted to the formulation of the main general theorems.
Specific families of bond market models are examined in Section
\ref{Specific models}. Proofs are postponed to Section \ref{Proofs
of the main theorems}.

\bigskip
\noindent {\bf Acknowledgement} The authors express thanks to
Professor D. Filipovi\'c for providing a copy of \cite{Morton} and a
section of a book to appear \cite{Filipovic}. The second author
thanks Professor S. Peszat for a useful discussion on the subject of
the paper.

\section{Preliminaries}\label{Preliminaries}
We fix here some notation and definitions needed in the sequel. We also
formulate our basic equation in a form easier to investigate. \vspace{3mm}

\noindent If $L$ is a L\'evy process with the Laplace transform
\begin{gather*}
\mathbf{E}(e^{-zL(t)})=e^{tJ(z)},\qquad t\in[0,T^\ast], \
z\in\mathbb{R},
\end{gather*}
then function $J$ is given by, see \cite{Bertoin}, \cite{Sato},
\cite{PeszatZabczyk},
\begin{gather}\label{Laplace transform}
J(z)=-az+\frac{1}{2}qz^2+\int_{\mathbb{R}}(e^{-zy}-1+zy\mathbf{1}_{(-1,
1)}(y)) \ \nu(dy),
\end{gather}
where $a\in\mathbb{R}, q\geq0$ and   $\nu$ is a measure which satisfies
integrability condition
\begin{gather}\label{war na miare Levyego}
\int_{\mathbb{R}}y^2\wedge \ 1 \ \nu(dy)<\infty.
\end{gather}
\vspace{3mm}

\noindent In this paper we examine the equation \eqref{f rownanie}
with noise being a purely discontinuous L\'evy process, without a
drift nor a Gaussian part. Thus $L$ is of the form
\begin{gather}\label{rownanie na szum}
L(t):=\int_{0}^{t}\int_{|y|<1} y \
\hat{\pi}(ds,dy)+\int_{0}^{t}\int_{|y|\geq 1}y \ {\pi}(ds,dy),
\end{gather}
where  $\pi$ is the Poisson random measure of jumps of $L$  and
$\hat{\pi}$ is the measure $\pi$ compensated by $dt \times \nu (dy)$.

Let us notice, that for each $T$ the solution  $f(t, T),\,\, t\in [0,T]$
of  \eqref{f rownanie} is a stochastic exponential and therefore (see
Theorem 37 in \cite{Protter}), equation \eqref{f rownanie} can be
equivalently written as:
\begin{multline}\label{basic}
\hspace*{1cm}f(t,T)=f_{0}(T)\,\, e^{\int_{0}^{t}J^{'}\left(\int_{s}^{T}\lambda(s,u)f(s-,u)du\right)\lambda(s,T)ds+\int_{0}^{t} \lambda (s, T) dL(s)}\\
   \hspace*{-1cm}   \cdot\prod_{s\leq
t}(1+\lambda (s,T)\triangle L(s))e^{-\lambda (s,T)\triangle L(s)},\qquad (t,T) \in \mathcal{T},
\end{multline}
where $\triangle L(s) = L(s) - L(s-)$. To limit our considerations
to models with non-negative forward rates, we impose  the following
natural assumptions. \bigskip

\noindent {\bf Standing assumptions:}
\begin{enumerate}[(K1)]
\item The initial curve $f_{0}$ is positive on $[0,T^\ast]$.
\item The support of the L\'evy measure is contained in the interval
    $(-1/\bar {\lambda},+\infty)\subseteq(-1,\infty)$.
\end{enumerate}

\noindent Under assumptions $(K1)$ and $(K2)$ we can write equation
(\ref{basic}) in the form
\begin{multline}\label{basic1}
f(t,T)=f_{0}(T)\,\,
e^{\int_{0}^{t}J^{'}\left(\int_{s}^{T}\lambda(s,u)f(s-,u)du\right)\lambda(s,T)ds+\int_{0}^{t}
\lambda (s, T)
dL(s)}\\[2ex]
      \cdot e^{\int_{0}^{t}\int_{-1/{\bar {\lambda}}}^{+\infty} \big(\ln (1+\lambda (s,T)y) -\lambda (s,T)y \big)\pi (ds, dy)},\qquad (t,T) \in
      \mathcal{T}.
\end{multline}
\noindent For brevity denote
\begin{gather}\label{wzor na a}
a(t,T):=f_{0}(T)
 e^{\int_{0}^{t} \lambda (s, T) dL(s) + \int_{0}^{t}\int_{-1/{\bar {\lambda}}}^{+\infty} \big(\ln (1+\lambda (s,T)y) -\lambda (s,T)y\big)\pi (ds, dy)}.
\end{gather}
Thus
\begin{equation}\label{basic2}
f(t,T) =
a(t,T)e^{\int_{0}^{t}J^{'}\left(\int_{s}^{T}\lambda(s,u)f(s-,u)du\right)\lambda(s,T)ds},\qquad
(t,T) \in
      \mathcal{T}.
\end{equation}
Since, for each $T$ the process, $\tilde{L}(t, T),\,\, t\in [0, T]$:
$$
\tilde{L}(t, T)= \int_{0}^{t} \lambda (s, T) dL(s) +
\int_{0}^{t}\int_{-1/{\bar {\lambda}}}^{+\infty} \Big(\ln (1+\lambda
(s,T)y) -\lambda (s,T)y\Big)\pi (ds, dy),\,\,t\in [0, T],
$$
has c\`adl\`ag trajectories
\begin{gather}\label{ogranicznie cadlag}
\sup_{t\in[0,T^\ast]}\tilde{L}(t,T)<\infty, \quad a.s.,
\end{gather}
and therefore $a(\cdot,T)$ is bounded on $[0,T]$ with probability
$1$. \bigskip

\bigskip
\noindent It turns out that due to the special form of the
coefficient $a$ given by \eqref{wzor na a} we can replace$f(s-,u)$
in \eqref{basic2} by $f(s,u)$.
\begin{prop}
Assume that $f$ is a bounded field. Then $f$ is a solution of
\eqref{basic2} if and only if
\begin{equation}
f(t,T) =
a(t,T)e^{\int_{0}^{t}J^{'}\left(\int_{s}^{T}\lambda(s,u)f(s,u)du\right)\lambda(s,T)ds},\qquad
(t,T) \in
      \mathcal{T}.
\end{equation}
\end{prop}
{\bf Proof:} We will show that for each $(t,T) \in \mathcal{T}$
\begin{gather*}
\int_{0}^{t}J^{'}\left(\int_{s}^{T}\lambda(s,u)f(s,u)du\right)\lambda(s,T)ds=
\int_{0}^{t}J^{'}\left(\int_{s}^{T}\lambda(s,u)f(s-,u)du\right)\lambda(s,T)ds.
\end{gather*}
Let us start with the observation that for $T\in[0,T^\ast]$ moments
of jumps of the process $f(\cdot,T)$ are the same as for
$a(\cdot,T)$. Moreover, it follows from \eqref{wzor na a} that the
set of jumps of $a(\cdot,T)$ is independent of $T$ and is contained
in the set
\begin{gather*}
\mathcal{Z}:=\{t\in[0,T^\ast]: \triangle L(t)\neq0\}.
\end{gather*}
Thus if $s\notin \mathcal{Z}$ then
\begin{gather*}
J^{'}\left(\int_{s}^{T}\lambda(s,u)f(s,u)du\right)\lambda(s,T)=
J^{'}\left(\int_{s}^{T}\lambda(s,u)f(s-,u)du\right)\lambda(s,T).
\end{gather*}
By Th. 2.8 in \cite{Apl} the set $\mathcal{Z}$ is at most countable,
so the assertion follows.\hfill$\square$

\noindent In the sequel we will examine equation \eqref{basic2} with
$f(s-,u)$ replaced by $f(s,u)$.

\subsection{Properties of J}
\noindent In virtue of \eqref{Laplace transform},  \eqref{rownanie na
szum} and the standing assumption (K2) the function $J$ is given by the
formula
\begin{align}\label{J u nas}\nonumber
J(z)&=\int_{\mathbb{R}}(e^{-zy}-1+zy\mathbf{1}_{(-1,1)}(y)) \ \nu(dy)\\[2ex]
&=\int_{-1/{\bar {\lambda}}}^{1}(e^{-zy}-1+zy) \ \nu(dy)+\int_{1}^{\infty}(e^{-zy}-1)
\ \nu(dy).
\end{align}
Taking into account \eqref{war na miare Levyego} we see that the
function $J$ is well defined for  $z\geq 0$. Let us notice that in
our setting we do not have to consider $J$ on the set $(-\infty,0)$.
Indeed, the assumptions $(K1)$ and $(K2)$ imply that $f$ is
positive, so the form of the equation \eqref{warunek HJM} together
with the condition \eqref{linearity} allow us to focus on the
properties of the function $J$ and its derivatives on the interval
$[0,\infty)$. Moreover, the condition \eqref{war na miare Levyego}
implies that for $z>0$ the function $J$ has derivatives of any order
and the following formulas hold, see Lemma 8.1 and 8.2 in
\cite{Rusinek},
\begin{align}\label{pierwsza pochodna J}
J^{'}(z)&=\int_{-1/{\bar {\lambda}}}^{1}y(1-e^{-zy})  \
\nu(dy)-\int_{1}^{\infty}ye^{-zy} \ \nu(dy),\quad J^{'}(0) = -\int_{1}^{\infty}y \nu (dy)\\[2ex]\label{druga pochodna
J} J^{''}(z)&=\int_{-1/{\bar {\lambda}}}^{\infty}y^2e^{-zy} \ \nu(dy),\quad
J^{'''}(z)=-\int_{-1/{\bar {\lambda}}}^{\infty}y^3e^{-zy} \ \nu(dy).
%\\[2ex]\label{trzecia pochodna
%J} .
\end{align}
\vspace{2mm}

\noindent Thus the objective of this paper is to examine existence
of a bounded solution for the equation
\begin{equation}\label{rownanie na f u nas}
f(t,T) =
a(t,T)e^{\int_{0}^{t}J^{'}\left(\int_{s}^{T}\lambda(s,u)f(s,u)du\right)\lambda(s,T)ds},\qquad
(t,T) \in
      \mathcal{T},
\end{equation}
where
$$
J^{'}(z)=\int_{-1/{\bar {\lambda}}}^{1}y(1-e^{-zy})\nu(dy)-\int_{1}^{\infty}ye^{-zy}\nu(dy)
,\quad z\geq 0,
$$
and the jump intensity measure $\nu$ is concentrated on $(-1/{\bar
{\lambda}}, 0)\cup(0,+\infty)$ and satisfies
\begin{gather}\label{condition on intensity}
\int_{(-1/{\bar {\lambda}}, 1)}y^2 \nu(dy) + \int_{1}^{\infty} y \nu (dy)
<\infty.
\end{gather}
Note that the function $J^{'}$ in the basic equation is increasing
on the whole interval $[0, + \infty)$ and $J^{'}(0)$ is either $0$,
if all jumps of $L$ are of size smaller or equal than $1$, or is
strictly negative. The latter integral in (\ref{condition on
intensity}) is required to be finite to imply that $J'(0)$ is
finite. Moreover, if
\begin{equation}\label{square integrability}
\int_{1}^{\infty} y^{2} \nu (dy)
<\infty ,
\end{equation}
then $J''$ is a bounded function on $[0, +\infty)$ and therefore
$J'$ is a Lipschitz function on $[0, +\infty)$. In fact,
(\ref{square integrability}) is also a necessary condition for $J'$
to be Lipschitz. The conditions (\ref{condition on intensity}) and
(\ref{square integrability}) are equivalent to the, respectively,
integrability and square integrability of the process $L$, see
\cite{Sato}.

\section{Main results}\label{Main results}
In this section we present formulation of the main theorems which
provide sufficient conditions for existence and non-existence
solution of the problem stated in Section \ref{Preliminaries}. Their
proofs are contained in Section \ref{Proofs of the main theorems}
and are preceded by a sequence of auxiliary results.

The following result provides sufficient conditions for existence of
a bounded solution.
\begin{tw}\label{tw 2 glowne}
Assume (\ref{condition on intensity}) and that
\begin{gather}\label{ujemne Jprim}
 \limsup_{z\rightarrow\infty} \ \left(\ln z-\bar {\lambda} T^\ast J^{'}(z)\right)=\infty.
\end{gather}
\begin{enumerate}[i)]
\item If the initial forward curve $f_{0}$ is bounded almost surely
then there exists a solution
$f:\mathcal{T}\longrightarrow\mathbb{R}_{+}$of \eqref{rownanie na f
u nas} which is also bounded almost surely.
\item If, in addition, (\ref{square integrability}) holds then the
solution $f$ is unique in the class of bounded fields.
\end{enumerate}
\end{tw}

\bigskip
The next results provide conditions which imply
non-existence of solution in the class of bounded fields and
explosions of locally bounded fields.
\begin{tw}\label{tw 1 glowne}
Assume (\ref{condition on intensity}), that $\lambda \equiv 1$ and for
some $\alpha>0$, $\beta\in\mathbb{R}$, $\gamma\in(0,1)$,
\begin{gather}\label{ogr na Jprim}
J^{'}(z)\geq\alpha z^\gamma+\beta, \qquad \forall z\geq0.
\end{gather}
For arbitrary $\kappa \in (0,1)$, there exists a positive constant
$K$ such that if
\begin{gather}\label{oddzielenie f_0}
f_{0}(T)> K, \quad\forall T\in[0,T^\ast],
\end{gather}
then there is no solution
$f:\mathcal{T}\longrightarrow\mathbb{R}_{+}$ of the equation
\eqref{rownanie na f u nas} which is bounded with probability
greater or equal than $\kappa$.
\end{tw}
\begin{tw}\label{tw1 o eksplozjach }
Assume that there exists a locally bounded solution of
\eqref{rownanie na f u nas} and that all the assumptions of Theorem
\ref{tw 1 glowne} are satisfied. Then
\begin{gather*}
\lim_{(t,T)\rightarrow(T^\ast,T^\ast)}f(t,T)=+\infty
\end{gather*} with probability
greater or equal than $\kappa$.
\end{tw}

\noindent In the case when $\lambda\equiv1$ and there is no solution
of equation \eqref{rownanie na f u nas} in the class of bounded
fields then one may ask if the solution does exist in a wider class
of fields satisfying some integrability conditions. However, in some
situations these two classes are the same. Assume, for example, that
the solution is supposed to satisfy condition:
\begin{gather*}
\int_{0}^{T^\ast}J^{'}\left(\int_{s}^{T^\ast}f(s,u)du\right)ds<\infty.
\end{gather*}
Then, due to the fact that $J^{'}(\cdot)$ is increasing, we see that
$f$ is well defined for any $(t,T)\in\mathcal{T}$. Moreover, if
$f_0$ is bounded, then for any $(t,T)\in\mathcal{T}$
\begin{align*}
f(t,T)&=e^{\int_{0}^{t}J^{'}\left(\int_{s}^{T}f(s,u)du\right)ds}\cdot
a(t,T)\\[2ex]
&\leq
e^{\int_{0}^{T^\ast}J^{'}\left(\int_{s}^{T^\ast}f(s,u)du\right)ds}\sup_{T\in[0,T^\ast]}f_0(T)\cdot\sup_{t\in[0,T^\ast]}
e^{L(t)+\int_{0}^{t}\int_{-1}^{\infty}\big(\ln(1+y)-y\big)\pi(ds,dy)}<\infty,
\end{align*}
and as a consequence $f$ is bounded.

\begin{rem}
Let the assumptions of Theorem \ref{tw 1 glowne} be satisfied and
that $f$ is a random field solving \eqref{rownanie na f u nas} and
for which
\begin{gather*}
\sup_{(t,T)\in\mathcal{T}_{x,y-\delta}}f(t,T)<\infty \quad
P-\text{a.s.},
\end{gather*}
for some $0<x\leq y\leq T^\ast$ and each $0<\delta<y$. Then
following the proof of Theorem \ref{tw1 o eksplozjach } one can show
that if $f_0$ is sufficiently large, then
\begin{gather*}
\lim_{(t,T)\uparrow(x,y)}f(t,T)=+\infty,
\end{gather*}
with probability arbitrarily close to $1$.
\end{rem}

\section{Specific models}\label{Specific models}
The crucial properties which imply existence or non-existence of
solution of the equation \eqref{rownanie na f u nas} are \eqref{ogr
na Jprim} and \eqref{ujemne Jprim}.  If \eqref{ogr na Jprim} holds
then there is no solution and if \eqref{ujemne Jprim} is satisfied
then there is a solution. It turns out that models with negative
jumps do not allow bounded solutions. For models with positive jumps
the answer does depend on the growth of the measure $\nu$ near $0$.
\subsection{Models with negative jumps}
\begin{tw}\label{tw 3}
If the measure $\nu$ has support in $(-1, 0)$ then the equation
\ref{rownanie na f u nas} with $\lambda\equiv1$ has no bounded
solutions.
\end{tw}
{\bf Proof:} Since
\begin{gather*}
J^{'''}(z)=-\int_{-1}^{0}y^3e^{-zy} \ \nu(dy)\geq 0, \qquad \forall
z\geq0,
\end{gather*}
the function $J'$ is convex and due to Lemma \ref{rem o funkcji wypuklej}
below the condition  \eqref{ogr na Jprim} is satisfied and it is enough to
apply Theorem \ref{tw 1 glowne} .\hfill$\square$

\begin{lem}\label{rem o funkcji wypuklej}
If $J^{'}$ is a convex function on $[0,\infty)$ then \eqref{ogr na Jprim}
is satisfied.
\end{lem}
{\bf Proof:} In virtue of the inequality $z\geq\sqrt{z}-1$, for $z\geq0$,
we have
\begin{gather*}
J^{'}(z)\geq J^{''}(0)z+J^{'}(0)\geq J^{''}(0)(\sqrt{z}-1)+J^{'}(0),
\qquad  \forall z\geq0.
\end{gather*}\hfill$\square$

\begin{tw}\label{przyklad 2}
Let $\nu$ be given by
\begin{gather*}
\nu(dy)=\frac{1}{\mid y\mid^{1+\rho}}\mathbf{1}_{(-1,1)}(y) \ dy, \,
\rho\in(0,2)\quad or\quad\nu(dy)=\frac{1}{\mid
y\mid^{1+\rho}}\mathbf{1}_{(-1,\infty)}(y) \ dy,\,\, \rho\in(1,2),
\end{gather*}
then  equation \eqref{rownanie na f u nas} with $\lambda\equiv1$ has
no bounded solutions.
\end{tw}
{\bf Proof:} We will show that
\begin{gather*}
J^{'}(z)\geq\frac{2}{2-\rho}\ z, \quad z\geq0.
\end{gather*}
in the first case and for some $\beta$,
\begin{gather*}
J^{'}(z)\geq \frac{2}{2-\rho}\ z-\beta, \quad z\geq0,
\end{gather*}
in the second case. By Theorem \ref{tw 1 glowne} the result will follow.

\noindent In virtue of \eqref{pierwsza pochodna J} we have
\begin{align*}
J^{'}(z)&=\int_{-1}^{1}y(1-e^{-zy})\frac{1}{\mid
y\mid^{1+\rho}}dy\\[2ex]
&=\int_{-1}^{0}y(1-e^{-zy})\frac{1}{(
-y)^{1+\rho}}dy+\int_{0}^{1}y(1-e^{-zy})\frac{1}{
y^{1+\rho}}dy\\[2ex]
&=-z^{\rho-1}\int_{0}^{z}\frac{1-e^v}{v^\rho}dv+z^{\rho-1}\int_{0}^{z}\frac{1-e^{-v}}{v^\rho}dv
=z^{\rho-1}\int_{0}^{z}\frac{e^v-e^{-v}}{v^\rho}dv.
\end{align*}
We use the series expansion
\begin{gather*}
e^{v}-e^{-v}=2\sum_{k=0}^{\infty}\frac{v^{2k+1}}{(2k+1)!}.
\end{gather*}
As a consequence we have
\begin{gather*}
\int_{0}^{z}\frac{e^{v}-e^{-v}}{v^{\rho}}dv=2\int_{0}^{z}\sum_{k=0}^{\infty}\frac{v^{2k+1-\rho}}{(2k+1)!}dv=2\sum_{k=0}^{\infty}
\frac{z^{2k+2-\rho}}{(2k+2-\rho)(2k+1)!}
\end{gather*}
and
\begin{gather*}
J^{'}(z)=z^{\rho-1}\int_{0}^{z}\frac{e^v-e^{-v}}{v^\rho}dv=2\sum_{k=0}^{\infty}
\frac{z^{2k+1}}{(2k+2-\rho)(2k+1)!}\geq\frac{2}{2-\rho}\ z.
\end{gather*}

\noindent We pass now to the second case.

\noindent Using \eqref{pierwsza pochodna J} and calculating as above we
have
\begin{align*}
J^{'}(z)&=z^{\rho-1}\int_{0}^{z}\frac{e^v-e^{-v}}{v^\rho}dv-\int_{1}^{\infty}\frac{e^{-zy}}{y^\rho}dy.
\end{align*}
For $z\geq0$ we have the following estimation
\begin{gather*}
\int_{1}^{\infty}\frac{e^{-zy}}{y^\rho}dy\leq\int_{1}^{\infty}\frac{1}{y^\rho}dy=:\beta<\infty
\end{gather*}
and as a consequence
\begin{gather*}
J^{'}(z)\geq \frac{2}{2-\rho}\ z-\beta, \quad z\geq0.
\end{gather*}
The proof is complete in virtue of the inequality $z\geq\sqrt{z}-1$
for $z\geq0$.\hfill$\square$
\begin{rem}\label{rho}
We restricted $\rho$ to the interval $(1,2)$ to satisfy (\ref{condition on
intensity}).
\end{rem}

\subsection{Models with positive jumps only}
We pass now to  models which generate bounded solutions and
therefore might be attractive for applications.

\noindent We start from the following theorem  which covers many
interesting cases with finite and infinite measure $\nu$.

\begin{tw}\label{przyklad6}
Let $\nu$ be a L\'evy measure on $(0,\infty)$ satisfying
\begin{gather*}
\int_{0}^{\infty}y \ \nu(dy)<\infty,
\end{gather*}
Then the equation \eqref{rownanie na f u nas} has a bounded
solution.
\end{tw}
{\bf Proof:} It is enough to prove that $J'$ is a bounded function. In
virtue of \eqref{pierwsza pochodna J} we have
\begin{gather*}
J^{'}(z)=\int_{0}^{1}y(1-e^{-zy})\nu(dy)-\int_{1}^{\infty}ye^{-zy}\nu(dy).
\end{gather*}
Since $J^{'}(z)\leq\int_{0}^{1}y \ \nu(dy)$ the boundedness follows.

\begin{tw}\label{przyklad4}
Let $\nu$ be given by
\begin{gather*}
\nu(dy)=\frac{1}{y^{1+\rho}}\mathbf{1}_{(0,1)}(y) \ dy, \qquad
\rho\in(0,2).
\end{gather*}
Then
\begin{enumerate}[1)]
\item if $\rho\in(1,2)$ then  equation \eqref{rownanie na f u nas} with $\lambda\equiv1$ has
    no bounded solutions
\item if $\rho\in(0,1)$  or
\item  $\rho=1$ and $\bar{\lambda}T^\ast<1$ then equation  \eqref{rownanie na f u
    nas} has a bounded solution.
\end{enumerate}
\end{tw}
{\bf Proof:} In virtue of \eqref{pierwsza pochodna J} we have
\begin{align}\label{Jprim w neg przykl 4}\nonumber
J^{'}(z)&=\int_{0}^{1}y(1-e^{-zy})\frac{1}{y^{1+\rho}} \ dy\\[2ex]
&=\int_{0}^{z}\frac{1-e^{-v}}{(\frac{v}{z})^\rho}\frac{1}{z} \
dv=z^{\rho-1}\int_{0}^{z}\frac{1-e^{-v}}{v^\rho} \ dv.
\end{align}
Let us consider the following cases.
\begin{enumerate}[1)]
\item $\rho\in(1,2)$\\
Then for $\alpha:=\int_{0}^{1}\frac{1-e^{-v}}{v^\rho}dv>0$ we have
\begin{gather*}
J^{'}(z)\geq\alpha z^{\rho-1} \quad \text{for} \ z\geq1.
\end{gather*}
The function $J^{'}$ is nonnegative on $[0,\infty)$ and thus
\begin{gather*}
J^{'}(z)\geq\alpha z^\gamma-\alpha \quad \text{for} \ z\geq 0,
\end{gather*}
with $\gamma:=\rho-1\in(0,1)$. As a consequence \eqref{ogr na Jprim}
is satisfied with $\beta=-\alpha$.
\item $\rho\in(0,1)$\\
We will show that $\lim_{z\rightarrow\infty} \bar{\lambda}T^\ast
J^{'}( z)<\infty$, what implies \eqref{ujemne Jprim}. We have
\begin{align*}
\lim_{z\rightarrow\infty} J^{'}(
z)&=\lim_{z\rightarrow\infty}  z^{\rho-1}\int_{0}^{z}\frac{1-e^{-v}}{v^\rho} \ dv\\[2ex]
&=\lim_{z\rightarrow\infty}\frac{\int_{0}^{z}\frac{1-e^{-v}}{v^\rho}
\ dv}{z^{1-\rho}}\overset{d'H}{=}\lim_{z\rightarrow\infty}
\frac{\frac{1-e^{-z}}{z^\rho}
}{(1-\rho)z^{-\rho}}\\[2ex]&=\lim_{z\rightarrow\infty}\frac{1-e^{-z}}{1-\rho}=\frac{1}{1-\rho}.
\end{align*}
\item $\rho=1$ and $\bar{\lambda}T^\ast<1$\\
One can check that in this case $J^{'}$ is unbounded and we can show
that
\begin{gather*}
\lim_{z\rightarrow\infty}\frac{\ln z}{\bar{\lambda}T^\ast
J^{'}(z)}>1.
\end{gather*}
This condition clearly implies $\eqref{ujemne Jprim}$. We have
\begin{align*}
\lim_{z\rightarrow\infty}\frac{\ln z}{\bar{\lambda}T^\ast
J^{'}(z)}\overset{d'H}{=}\lim_{z\rightarrow\infty}\frac{\frac{1}{z}}{\frac{1-e^{-z}}{z}\cdot{\bar{\lambda}T^\ast}}
=\lim_{z\rightarrow\infty}\frac{1}{\bar{\lambda}T^\ast(1-e^{-z})}=\frac{1}{\bar{\lambda}T^\ast}>1.
\end{align*}
\hfill$\square$
\end{enumerate}
Our final class of examples is with large jumps.
\begin{tw}\label{large jumps}
Let $\nu$ be given by
\begin{gather*}
\nu(dy)=\frac{1}{y^{1+\rho}}\mathbf{1}_{(0,\infty)}(y) \ dy, \qquad
\rho\in(1,2).
\end{gather*}
Then the equation \eqref{rownanie na f u nas} with $\lambda\equiv1$
has no bounded solutions.
\end{tw}
{\bf Proof:} In virtue of \eqref{pierwsza pochodna J} we have
\begin{align}\label{Jprim w neg przykl 5}\nonumber
J^{'}(z)&=\int_{0}^{1}y(1-e^{-zy})\frac{1}{y^{1+\rho}}dy-\int_{1}^{\infty}ye^{-zy}\frac{1}{y^{1+\rho}}dy\\[2ex]
&=z^{\rho-1}\int_{0}^{z}\frac{1-e^{-v}}{v^\rho}dv-\int_{1}^{\infty}\frac{e^{-zy}}{y^\rho}dy.
\end{align}
Due to the inequality
\begin{gather*}
\int_{1}^{\infty}\frac{e^{-zy}}{y^\rho}dy\leq\int_{1}^{\infty}\frac{1}{y^\rho}dy<\infty,\quad
z\geq0,
\end{gather*}
and the estimation from the proof of Th. \ref{przyklad4}, $(1)$ we
have
\begin{gather*}
J^{'}(z)\geq\alpha
z^\gamma-\alpha-\int_{1}^{\infty}\frac{1}{y^\rho}dy,\quad z\geq0,
\end{gather*}
so \eqref{ogr na Jprim} holds with
$\alpha=\int_{0}^{1}\frac{1-e^{-v}}{v^\rho}dv$, $\gamma=\rho-1$,
$\beta=-\alpha-\int_{1}^{\infty}\frac{1}{y^\rho}dy$. \hfill$\square$

\section{Proofs of the main theorems}\label{Proofs of the main theorems}
This section is divided into two parts containing proofs of Theorems
\ref{tw 1 glowne}, \ref{tw1 o eksplozjach } and \ref{tw 2 glowne}
respectively with all auxiliary lemmas and propositions.

\subsection{Non-existence}
Recall that the sets $\mathcal{T}$ and $\mathcal{T}_{x,y}$, where
$0<x\leq T^\ast$, $0<y\leq T^\ast$ are given by \eqref{dziedzina f}
and \eqref{def zbioru T_x,y}. In the sequel we will use the
notation: $\bar{\mathbb{R}}_{+}:=\mathbb{R}_{+}\cup\{+\infty\}$.

\begin{lem}\label{lem o szacowaniu całek}
Let $f:[a,b]\longrightarrow\mathbb{R}_{+}$, where
$a,b\in\mathbb{R},a<b$, be a continuous function. For any
$\gamma\in(0,1)$ we have
\begin{gather}\label{nierownosc dla calek}
\int_{a}^{b}f^\gamma(x)dx\leq(b-a)^{1-\gamma}\left(\int_{a}^{b}f(x)dx\right)^{\gamma}.
\end{gather}
\end{lem}
{\bf Proof:}   If $z_1,z_2,...,z_n$ are positive reals and $\gamma \in
(0,1)$  then
\begin{gather}\label{nierownosc dla srednich uogolnionych}
\left(\frac{1}{n}\sum_{i=1}^{n}z_i^{\gamma}\right)^{\frac{1}{\gamma}}\leq\left(\frac{1}{n}\sum_{i=1}^{n}z_i\right).
\end{gather}
In fact, by H\"older inequality with $p={\frac{1}{\gamma}}$ and
$q={\frac{1}{1 - \gamma}}$,
$$
\sum_{i=1}^{n}z_i^{\gamma}\leq
\left(\sum_{i=1}^{n}(z_i^{\gamma})^{{\frac{1}{\gamma}}}\right)^{\gamma}
\left(\sum_{i=1}^{n} 1^{{\frac{1}{1 - \gamma}}}\right)^{1-\gamma} ,
$$
and rearranging terms one gets (\ref{nierownosc dla srednich
uogolnionych}).

\noindent Let us consider an equidistant partition of the interval $[a,b]$
with $x_i=a+i\cdot\frac{b-a}{n}$, $i=1,2,...,n$. Using \eqref{nierownosc
dla srednich uogolnionych} with $z_i=f(x_i), i=1,2,...,n$,   we obtain
\begin{align}\label{nierownosc dla sum}\nonumber
\sum_{i=1}^{n}\frac{b-a}{n}f^\gamma(x_i)=(b-a)\left(\frac{1}{n}\sum_{i=1}^{n}f^\gamma(x_i)\right)&\leq(b-a)\left(\frac{1}{n}\sum_{i=1}^{n}f(x_i)\right)^\gamma\\[2ex]
&=(b-a)^{1-\gamma}\left(\sum_{i=1}^{n}\frac{b-a}{n}f(x_i)\right)^\gamma
\end{align}
Letting $n\longrightarrow\infty$ in \eqref{nierownosc dla sum} we obtain
\eqref{nierownosc dla calek}.\hfill$\square$ \vspace{3mm}

\noindent In the following, for any $\alpha>0$ and $\gamma\in(0,1)$
and $0<x\leq y\leq T^\ast$, we will consider the function
$h:\mathcal{T}_{x,y}\longrightarrow\bar{\mathbb{R}}_{+}$ given by
\begin{equation}\label{funkcja wybuchowa}
h(t,T):= \left\{\begin{array}{ll}
\left(\frac{1}{x-t+y-T}\right)^{\frac{3}{\gamma}}&\text{for}
\ (t,T)\neq(x,y)\\[2ex]
\infty&\text{for} \ (t,T)=(x,y),
\end{array}\right.
\end{equation}
and the function $R:\mathbb{R}_{+}\longrightarrow\mathbb{R}_{+}$
defined as
\begin{gather}\label{funkcja R}
R(z):=\alpha z\mathbf{1}_{[0,1]}(z)+\alpha
z^{\gamma}\mathbf{1}_{(1,\infty)}(z)\qquad z\in\mathbb{R}_{+}.
\end{gather}
The following properties of the function $R$ can be easily verified
\begin{align}\label{pierwsza wlasnosc R}
\alpha z^{\gamma}\geq R(z)&\geq \alpha z^{\gamma}-1, \qquad
z\in\mathbb{R}_{+},\\[2ex] \label{druga wlasnosc R} |R(z_1)-R(z_2)|&\leq
\alpha|z_1-z_2|, \qquad z_1,z_2\in\mathbb{R}_{+}.
\end{align}

\begin{prop}\label{prop ciaglosc g}
Let $\alpha>0$, $\gamma\in(0,1)$, $0<x\leq y\leq T^\ast$ and the
functions $h, R$ be given by \eqref{funkcja wybuchowa} and
\eqref{funkcja R} respectively. The function
$g:\mathcal{T}_{x,y}\longrightarrow\mathbb{R}_{+}$ defined by the
formula
\begin{equation}\label{mniejsza funkcja}
g(t,T):= \left\{\begin{array}{ll}
e^{-\int_{0}^{t}R\left(\int_{s}^{T}h(s,u)du\right) ds}\cdot
h(t,T)&\emph{for}
\ (t,T)\neq(x,y)\\[2ex]
0&\emph{for} \ (t,T)=(x,y)
\end{array}\right.
\end{equation}
is continuous.
\end{prop}
{\bf Proof:} Let us start with an auxiliary calculation and
estimation. One can check that
\begin{gather}\label{wyliczona całka}
\int_{0}^{t}\int_{s}^{T}\frac{1}{(x-s+y-u)^3}\ du \
ds=\frac{t}{2}\cdot\frac{-T^2-Tt-ty+2Ty+2Tx-tx}{(x-t+y-T)(x+y-2t)(x+y-T)(x+y)},
\end{gather}
for any $(t,T)\in\mathcal{T}_{x,y}$.

 \noindent In virtue of Lemma
\ref{lem o szacowaniu całek} we have
\begin{align}\label{oszacowanie wykładnika}\nonumber
\int_{0}^{t}\left(\int_{s}^{T}h(s,u)du\right)^\gamma ds&\geq
\int_{0}^{t}\left((T-s)^{\gamma-1}\int_{s}^{T}h^\gamma(s,u)du\right)ds\\[2ex]\nonumber
&\geq
T^{\gamma-1}\int_{0}^{t}\left(\int_{s}^{T}h^\gamma(s,u)du\right)ds\\[2ex]
&\geq T^{\gamma-1}\int_{0}^{t}\int_{s}^{T}\frac{1}{(x-s+y-u)^3} \
duds.
\end{align}
As a consequence of \eqref{pierwsza wlasnosc R}, \eqref{oszacowanie
wykładnika} and \eqref{wyliczona całka} we have
\begin{align*}
e^{-\int_{0}^{t}R\left(\int_{s}^{T}h(s,u)du\right) ds}\cdot h(t,T)
&\leq
e^{-\int_{0}^{t}\left\{\alpha\left(\int_{s}^{T}h(s,u)du\right)^{\gamma}-1\right\}
ds}\cdot h(t,T)\\[2ex]
&=e^{-\alpha\int_{0}^{t}\left(\int_{s}^{T}h(s,u)du\right)^{\gamma}
ds}\cdot e^{t}\cdot h(t,T)\\[2ex]
&\leq e^{-\alpha
T^{\gamma-1}\int_{0}^{t}\int_{s}^{T}\frac{1}{(x-s+y-u)^3} \
duds}\cdot e^t\cdot
\left(\frac{1}{x-t+y-T}\right)^{\frac{3}{\gamma}}\\[2ex]
&\leq e^{-\frac{\alpha t
T^{\gamma-1}}{2}\cdot\frac{-T^2-Tt-ty+2Ty+2Tx-tx}{(x-t+y-T)(x+y-2t)(x+y-T)(x+y)}}\cdot
e^t\cdot \left(\frac{1}{x-t+y-T}\right)^{\frac{3}{\gamma}}.
\end{align*}
We need to show continuity of $g$ only in the point $(x,y)$. We have
\begin{align*}
\lim_{t\rightarrow x, T\rightarrow
y}(-T^2-Tt-ty+2Ty+2Tx-tx)&=y^2-x^2>0\\[1ex]
\lim_{t\rightarrow x, T\rightarrow y}(x+y-2t)&=y-x>0\\[1ex]
\lim_{t\rightarrow x, T\rightarrow y}(x+y-T)&=x>0.
\end{align*}
Thus to show that
\begin{gather*}
\lim_{t\rightarrow x, T\rightarrow y}g(t,T)=0
\end{gather*}
it is enough to notice that
\begin{gather*}
\lim_{z\rightarrow\infty}e^{-cz}z^{\frac{3}{\gamma}}=0,
\qquad\text{for} \ c>0.
\end{gather*}
\hfill$\square$

\begin{rem}\label{prop o zwiazku g z h}
Let $\alpha>0$, $\gamma\in(0,1)$, $0<x\leq y\leq T^\ast$. The
functions $h,R,g$ given by \eqref{funkcja wybuchowa}, \eqref{funkcja
R}, \eqref{mniejsza funkcja} satisfy the following equation
\begin{gather*}
h(t,T)=e^{\int_{0}^{t}R\left(\int_{s}^{T}h(s,u)du\right)ds}\cdot
g(t,T),\qquad \forall (t,T)\in\mathcal{T}_{x,y}.
\end{gather*}
\end{rem}
{\bf Proof:} We have
\begin{align*}
h(t,T)&=e^{\int_{0}^{t}R\left(\int_{s}^{T}h(s,u)du\right) ds}\cdot
e^{-\int_{0}^{t}R\left(\int_{s}^{T}h(s,u)du\right)
ds}\cdot h(t,T)\\[2ex]
&=e^{\int_{0}^{t}R\left(\int_{s}^{T}h(s,u)du\right) ds}\cdot
g(t,T),\qquad \forall (t,T)\in\mathcal{T}_{x,y}.
\end{align*}
\hfill$\square$

\begin{lem}\label{prop Gronwall}
Let $0<t_0\leq T_0<\infty$ and define a set
\begin{gather*}
A:=\Big\{(t,T): t\leq T, \ 0\leq t\leq t_0, \ t\leq T\leq T_0\Big\}.
\end{gather*}
If $d:A\longrightarrow\mathbb{R}_{+}$ is a bounded function
satisfying
\begin{gather}\label{rownanie w Gronwallu}
d(t,T)\leq K \int_{0}^{t}\int_{s}^{T}d(s,u)duds \qquad\forall
(t,T)\in A
\end{gather}
where $0<K<\infty$ then $d(t,T)\equiv0$ on $A$.
\end{lem}
{\bf Proof:} Assume that $d$ is bounded by a constant $M>0$ on $A$.
We show inductively that
\begin{gather}\label{nierownosc indukcyjna}
d(t,T)\leq MK^{n}\frac{(tT)^n}{(n!)^2}, \qquad\forall(t,T)\in A.
\end{gather}
The formula $\eqref{nierownosc indukcyjna}$ is valid for $n=0$.
Assume that it is true for some $n$ and show that it is true for
$n+1$. We have the following estimation
\begin{align*}
d(t,T)&\leq K\int_{0}^{t}\int_{s}^{T}MK^{n}\frac{(su)^n}{(n!)^2}duds=MK^{n+1}\frac{1}{(n!)^2}\int_{0}^{t}s^n(\int_{s}^{T}u^ndu)ds\\[2ex]
&=
MK^{n+1}\frac{1}{(n!)^2}\int_{0}^{t}s^n\left(\frac{T^{n+1}-s^{n+1}}{n+1}\right)ds\leq
MK^{n+1}\frac{1}{(n!)^2}\int_{0}^{t}s^n\frac{T^{n+1}}{n+1}ds\\[2ex]
&=MK^{n+1}\frac{1}{(n!)^2}\frac{t^{n+1}}{(n+1)}\frac{T^{n+1}}{(n+1)}=MK^{n+1}\frac{(tT)^{n+1}}{((n+1)!)^2}.
\end{align*}
Letting $n\longrightarrow\infty$ in $\eqref{nierownosc indukcyjna}$
we see that $d(t,T)=0$. \hfill$\square$
\begin{prop}\label{prop jedynosc rozwiazania h}
Let $0<x\leq y\leq T^\ast$, $0<\delta<y$ and
$g:\mathcal{T}_{x,y-\delta}\longrightarrow\mathbb{R}_{+}$ be a
bounded function. Assume that there exists a bounded function
$h:\mathcal{T}_{x,y-\delta}\longrightarrow\mathbb{R}_{+}$ which
solves the following equation
\begin{gather}\label{rownanie w prop. o jedynosci}
h(t,T)=e^{\int_{0}^{t}R\left(\int_{s}^{T}h(s,u)du\right) ds}\cdot
g(t,T),\qquad \forall (t,T)\in\mathcal{T}_{x,y-\delta},
\end{gather}
where $R$ is given by \eqref{funkcja R}. Then $h$ is uniquely
determined in the class of bounded functions on
$\mathcal{T}_{x,y-\delta}$.
\end{prop}
{\bf Proof:} Assume that
$h_1,h_2:\mathcal{T}_{x,y-\delta}\longrightarrow\mathbb{R}_{+}$ are
bounded solutions of \eqref{rownanie w prop. o jedynosci}. Then the
function $\mid h_1-h_2\mid$ is bounded and satisfies
\begin{gather*}
\mid h_1(t,T)-h_2(t,T)\mid\leq \parallel g\parallel\cdot\mid
e^{\int_{0}^{t}R\left(\int_{s}^{T}h_1(s,u)du\right)
ds}-e^{\int_{0}^{t}R\left(\int_{s}^{T}h_2(s,u)du\right) ds}\mid,
\qquad \forall (t,T)\in\mathcal{T}_{x,y-\delta},
\end{gather*}
where
\begin{gather*}
\parallel g\parallel=\sup_{(t,T)\in\mathcal{T}_{x,y-\delta}}\mid g(t,T)\mid.
\end{gather*}
As a consequence of the inequality $\mid
e^{x}-e^{y}\mid\leq\max\{e^x,e^y\}\mid x-y\mid$ for
$x,y\in\mathbb{R}$ we have
\begin{gather*}
\mid h_1(t,T)-h_2(t,T)\mid\leq K \int_{0}^{t}\left|
R\left(\int_{s}^{T}
h_1(s,u)du\right)-R\left(\int_{s}^{T}h_2(s,u)du\right)\right|
ds,\qquad \forall (t,T)\in\mathcal{T}_{x,y-\delta},
\end{gather*}
where
\begin{gather*}
K:=\parallel g\parallel
\sup_{(t,T)\in\mathcal{T}_{x,y-\delta}}\max_{i=1,2}\Big\{e^{\int_{0}^{t}R\left(\int_{s}^{T}h_i(s,u)du\right)
ds}\Big\}<\infty.
\end{gather*}
In virtue of \eqref{druga wlasnosc R} we have
\begin{gather*}
\mid h_1(t,T)-h_2(t,T)\mid\leq \alpha K \int_{0}^{t}\int_{s}^{T}\mid
h_1(s,u)-h_2(s,u)\mid duds,\qquad \forall
(t,T)\in\mathcal{T}_{x,y-\delta}.
\end{gather*}
In view of Lemma \ref{prop Gronwall}, with $t_0=\min\{x,y-\delta\}$,
$T_{0}=y-\delta$, we have $h_1(t,T)=h_2(t,T)$ for all
$(t,T)\in\mathcal{T}_{x,y-\delta}$. \phantom{a}\hfill$\square$

\begin{prop}\label{prop f>h}
Let $\alpha>0$, $\gamma\in(0,1)$ and function $R$ be given by
\eqref{funkcja R}. Let
$f_{1}:\mathcal{T}_{x,y-\delta}\longrightarrow\mathbb{R}_{+}$, where
$0<x\leq y\leq T^\ast$; $0<\delta<y-x$, be a bounded function
satisfying inequality
\begin{gather}\label{nierwnosc dla f}
f_1(t,T)\geq e^{\int_{0}^{t}R\left(\int_{s}^{T}f_1(s,u)du\right)
ds}\cdot g_1(t,T), \qquad \forall (t,T)\in\mathcal{T}_{x,y-\delta},
\end{gather}
where $g_1:\mathcal{T}_{x,y-\delta}\longrightarrow\mathbb{R}_{+}$.
Let $f_2:\mathcal{T}_{x,y-\delta}\longrightarrow\mathbb{R}_{+}$ be a
bounded function solving equation
\begin{gather}
f_2(t,T)= e^{\int_{0}^{t}R\left(\int_{s}^{T}f_2(s,u)du\right)
ds}\cdot g_2(t,T), \qquad \forall (t,T)\in\mathcal{T}_{x,y-\delta},
\end{gather}
where $g_2:\mathcal{T}_{x,y-\delta}\longrightarrow\mathbb{R}_{+}$ is
a bounded function. Moreover, assume that
\begin{gather}\label{g_1>g_2}
g_1(t,T)\geq g_2(t,T)\geq 0, \qquad
\forall(t,T)\in\mathcal{T}_{x,y-\delta}.
\end{gather}
Then $f_1(t,T)\geq f_2(t,T)$ for all
$(t,T)\in\mathcal{T}_{x,y-\delta}$.
\end{prop}
{\bf Proof:} Let us define the operator $\mathcal{K}$ acting on
bounded functions on $\mathcal{T}_{x,y-\delta}$ by
\begin{gather}\label{def operatora K}
\mathcal{K}k(t,T):=e^{\int_{0}^{t}R\left(\int_{s}^{T}k(s,u)du\right)
ds}\cdot g_2(t,T), \qquad (t,T)\in\mathcal{T}_{x,y-\delta}.
\end{gather}
Let us notice that in view of \eqref{nierwnosc dla
f},\eqref{g_1>g_2} and \eqref{def operatora K} we have
\begin{gather}\label{Kf<f}
\mathcal{K}f_1(t,T)\leq
e^{\int_{0}^{t}R\left(\int_{s}^{T}f_1(s,u)du\right) ds}\cdot
g_1(t,T)\leq f_1(t,T),
\qquad\forall(t,T)\in\mathcal{T}_{x,y-\delta}.
\end{gather}
It is clear that the operator $\mathcal{K}$ is monotonic, i.e.
\begin{gather}\label{monotonicznosc K}
k_1(t,T)\leq k_2(t,T)\quad\forall(t,T)\in\mathcal{T}_{x,y-\delta}
\quad\Longrightarrow\quad
\mathcal{K}k_1(t,T)\leq\mathcal{K}k_2(t,T)\quad\forall(t,T)\in\mathcal{T}_{x,y-\delta}.
\end{gather}
Let us consider the sequence of functions:
$f_1,\mathcal{K}f_1,\mathcal{K}^2f_1$,... . In virtue of \eqref{Kf<f} and
\eqref{monotonicznosc K} we see that
$f_1\geq\mathcal{K}f_1\geq\mathcal{K}^2f_1\geq$... .Thus this sequence is
pointwise convergent to some function $\bar{f}$ and it is bounded by
$f_1$, so applying the dominated convergence theorem in the formula
\begin{gather*}
\mathcal{K}^{n+1}f_1(t,T)=e^{\int_{0}^{t}R\left(\int_{s}^{T}\mathcal{K}^nf_1(s,u)du\right)
ds}\cdot g_2(t,T),\qquad \forall(t,T)\in\mathcal{T}_{x,y-\delta}
\end{gather*}
we obtain
\begin{gather*}
\bar{f}(t,T)=e^{\int_{0}^{t}R\left(\int_{s}^{T}\bar{f}(s,u)du\right)
ds}\cdot g_2(t,T),\qquad \forall(t,T)\in\mathcal{T}_{x,y-\delta}.
\end{gather*}

\noindent Moreover, $\bar{f}$ is bounded and thus, in view of
Proposition \ref{prop jedynosc rozwiazania h}, we have
$\bar{f}=f_2$. As a consequence $f_1\geq f_2$ on
$\mathcal{T}_{x,y-\delta}$. \hfill $\square$

\bigskip
\noindent
{\bf Proof of Theorem \ref{tw 1 glowne}}\\
\noindent Assume that there exists a bounded solution of
\eqref{rownanie na f u nas}. Fix any $(x,y)\in\mathcal{T}$ such that
$x>0$ and three deterministic functions
$h:\mathcal{T}_{x,y}\longrightarrow\bar{\mathbb{R}}_{+}$,
$R:\mathbb{R}_{+}\longrightarrow\mathbb{R}_{+}$,
$g:\mathcal{T}_{x,y}\longrightarrow\mathbb{R}_{+}$ given by
\eqref{funkcja wybuchowa}, \eqref{funkcja R} and \eqref{mniejsza
funkcja} respectively. Recall that, due to Remark \ref{prop o
zwiazku g z h}, they satisfy the equation
\begin{gather}\label{h i g}
h(t,T)=e^{\int_{0}^{t}R\left(\int_{s}^{T}h(s,u)du\right) ds}\cdot
g(t,T),\qquad \forall (t,T)\in\mathcal{T}_{x,y}.
\end{gather}
\noindent Due to \eqref{ogr na Jprim} and \eqref{pierwsza wlasnosc R}, the
forward rate $f$ satisfies the following inequality
\begin{align}\label{nierownosc dla tilde f }\nonumber
f(t,T)&=e^{\int_{0}^{t}J^{'}(\int_{s}^{T} f (s,u)du)ds} a(t,T)\\[2ex]\nonumber
&\geq
e^{\int_{0}^{t}\alpha\left(\int_{s}^{T}f(s,u)du\right)^\gamma
ds}e^{\beta t} a(t,T)\\[2ex]
&\geq e^{\int_{0}^{t}R\left(\int_{s}^{T}f (s,u)du\right)
ds}e^{\beta t} a(t,T),\qquad\forall(t,T)\in\mathcal{T}.
\end{align}
In virtue of Proposition \ref{prop ciaglosc g} the function $g$ is
continuous on $\mathcal{T}_{x,y}$ and thus bounded. Thus, see
(\ref{ogranicznie cadlag}), if the constant $K$ is sufficiently large,
with a probability arbitrarily close to $1$,
\begin{equation}\label{initial condition}
e^{\beta t} a(t,T) \geq g(t,T),\qquad\forall(t,T)\in\mathcal{T}_{x,y}.
\end{equation}
\noindent Let us fix $0<\delta<y$ and consider inequality
\eqref{nierownosc dla tilde f } and equality \eqref{h i g} on the
set $\mathcal{T}_{x,y-\delta}$. Then the function $h$ is continuous.
In virtue of Proposition \ref{prop f>h} we have
\begin{gather*}
f(t,T)\geq h(t,T), \qquad\forall(t,T)\in\mathcal{T}_{x,y-\delta}.
\end{gather*}
As a consequence we have
\begin{align*}
f(t,T)\geq
h(t,T)=\frac{1}{(x-t+y-T)^{\frac{3}{\gamma}}},\qquad\forall(t,T)\in\mathcal{T}_{x,y-\delta}.
\end{align*}
For any sequence $(t_n,T_n)\in\mathcal{T}_{x,y}$ satisfying $t_n\uparrow
x$, $T_n\uparrow y$ define a sequence $\delta_n:=\frac{y-T_n}{2}$. Then
\begin{gather*}
f(t,T)\geq\frac{c}{(x-t+y-T)^{\frac{3}{\gamma}}},\qquad\forall(t,T)\in\mathcal{T}_{x,y-\delta_n},
\end{gather*}
and in particular
$f(t_n,T_n)\geq\frac{c}{(x-t_n+y-T_n)^{\frac{3}{\gamma}}}$. As a
consequence $\lim_{n\rightarrow\infty}f(t_n,T_n)=+\infty$ what is a
contradiction with the assumption that $f$ is bounded.
\hfill$\square$

\bigskip
\noindent The proof of Th.\ref{tw1 o eksplozjach } can be deduced
from the proof of Th.\ref{tw 1 glowne}.

\medskip
\noindent
{\bf Proof of Theorem \ref{tw1 o eksplozjach }}\\
We follow the proof of Th.\ref{tw 1 glowne} with $x=T^\ast$,
$y=T^\ast$. From the fact that $f$ is locally bounded we have
\begin{gather*}
f(t,T)\geq h(t,T),
\qquad\forall(t,T)\in\mathcal{T}_{T^\ast,T^\ast-\delta},
\end{gather*}
for each $0<\delta<T^\ast$. As a consequence
\begin{gather*}
\lim_{(t,T)\rightarrow(T^\ast,T^\ast)}f(t,T)=+\infty.
\end{gather*}
\hfill $\square$

\subsection{Existence} We can write \eqref{rownanie na f u nas} in
the form $f=\mathcal{A}f$, where
\begin{gather}\label{def operatora A}
\mathcal{A}h(t,T):=a(t,T)\cdot
e^{\int_{0}^{t}J^{'}\big(\int_{s}^{T}\lambda (s,u)
h(s,u)du\big)\lambda (s,T) ds},\qquad (t,T)\in \mathcal{T}.
\end{gather}
The proof of Theorem \ref{tw 2 glowne} is based on the properties of
the operator $\mathcal{A}$. If we fix $\omega\in\Omega$ then we can
treat $\mathcal{A}$ as a purely deterministic transformation with
the function $a$ positive and bounded.
\begin{prop}\label{prop o stalej c dla operatora A}
Assume that the function $J^{'}$ satisfies  \eqref{ujemne Jprim} and
$a$ is a nonnegative function bounded from above by some constant
$K$. Then there exists a positive constant $c$ such that if
\begin{gather*}
h(t,T)\leq c, \quad \forall (t,T)\in\mathcal{T}
\end{gather*}
 for a non-negative function $h$, then
\begin{gather}\label{warunek ograniczajacy na Ah}
\mathcal{A}h(t,T)\leq c, \quad \forall (t,T)\in\mathcal{T}.
\end{gather}
\end{prop}
{\bf Proof:} Let us assume that $h(t,T)\leq c$ for all
$(t,T)\in\mathcal{T}$ for some positive c.  Using the fact that
$J^{'}$ is increasing and $\lambda$ positive, we have
\begin{equation}\label{wyjsciowe szacowanie Ah}\nonumber
\mathcal{A}h(t,T) \leq a(t,T)\cdot
e^{J^{'}(\bar {\lambda} c T^{\ast})  \int_{0}^{t} \lambda (s,T)ds}
\end{equation}
Since $a$ is bounded by a constant $K$ we arrive at the following
inequality
$$
\mathcal{A}h(t,T) \leq K e^{J^{'}(\bar {\lambda} c T^\ast) \int_{0}^{t} \lambda (s,T)ds },\,\,\,(t,T)\in\mathcal{T}.
$$
It is therefore enough to find a positive constant $c$ such that
\begin{gather}\label{nierownosc logarytmowa}
\ln K + J^{'}(\bar {\lambda} c T^\ast)\cdot \int_{0}^{t} \lambda (s,T)ds
\leq \ln c,\,\,(t,T)\in\mathcal{T}.
\end{gather}
If the function $J^{'}$ is negative on $[0, +\infty)$ then it is
enough to take $c=K$. If $J^{'}$ takes positive values then it is
enough to find a positive an arbitrarily large constant $c$ such
that
\begin{gather}\label{nierownosc logarytmowa}
\ln K + \bar {\lambda} T^\ast \cdot J^{'}(\bar {\lambda} c T^\ast)\leq \ln
c,\,\,(t,T)\in\mathcal{T}.
\end{gather}
Existence of such $c$ is an immediate  consequence of the assumption
(\ref{ujemne Jprim}).\hfill$\square$

\bigskip
\noindent
{\bf Proof of Theorem \ref{tw 2 glowne}}\\
\noindent Part $i)$. The operator $\mathcal{A}$ is monotonic, i.e.
\begin{gather*}
h_1\leq h_2 \quad\Longrightarrow\quad \mathcal{A}h_1\leq
\mathcal{A}h_2.
\end{gather*}
The sequence $h_0\equiv 0, \ h_{n+1}:=\mathcal{A}h_n$ is thus
monotonically increasing to $\bar{h}$ and by the monotone
convergence theorem we have
\begin{gather*}
\bar{h}(t,T)=\mathcal{A}\bar{h}(t,T), \qquad
\forall(t,T)\in\mathcal{T}.
\end{gather*}
Moreover, since $h_0\leq c$, where $c=c(\omega)$ is given by
Proposition \ref{prop o stalej c dla operatora A}, $\bar{h}$ is
bounded. From the form of the operator $\mathcal{A}$ it follows that
$\bar{h}(\cdot,T)$ is c\`adl\`ag for each $T\in[0,T^\ast]$.
Conditions \eqref{1war na rozw} and \eqref{3war na rozw} follows
from the fact that $\bar{h}$ is a pointwise limit. \vspace{2mm}

\noindent Part $ii)$. The function $J'$ is Lipschitz on $[0,
+\infty)$ and therefore we can repeat all arguments from the proof
of Proposition \ref{prop jedynosc rozwiazania h} and the result
follows. \hfill$\square$

\end{document}